\documentclass[journal]{IEEEtran}
\usepackage{cite}

\usepackage{braket}

\ifCLASSINFOpdf
\else
\fi
\usepackage{amsmath}
\usepackage{mathtools}
\usepackage{amssymb}
\usepackage{dsfont}
\usepackage{algorithmic}
\usepackage{url}

\usepackage{mathtools}
\usepackage{makecell}

\begin{document}
%
\title{Data-driven Modality Fusion: An AI-enabled Framework for Large-Scale Sensor Network Management}

%
%
%

\author{Hrishikesh Dutta,~\IEEEmembership{Member,~IEEE,}
        Roberto Minerva,~\IEEEmembership{Senior Member,~IEEE,}
        Maira Alvi,~\IEEEmembership{Student Member,~IEEE,}
        and~Noel Crespi,~\IEEEmembership{Senior Member,~IEEE}
\thanks{The authors are with the Data Intelligence and Communication Engineering Lab, Telecom SudParis, Institut Polytechnique de Paris, France (email: hrishikesh.dutta@telecom-sudparis.eu; roberto.minerva@telecom-sudparis.eu; maira.alvi@telecom-sudparis.eu;  noel.crespi@telecom-sudparis.eu)}
}

%
%

\maketitle

\begin{abstract}
The development and operation of smart cities rely heavily on large-scale Internet-of-Things (IoT) networks and sensor infrastructures that continuously monitor various aspects of urban environments. These networks generate vast amounts of data, posing challenges related to bandwidth usage, energy consumption, and system scalability. This paper introduces a novel sensing paradigm called Data-driven Modality Fusion (DMF), designed to enhance the efficiency of smart city IoT network management. By leveraging correlations between time-series data from different sensing modalities, the proposed DMF approach reduces the number of physical sensors required for monitoring, thereby minimizing energy expenditure, communication bandwidth, and overall deployment costs. The framework relocates computational complexity from the edge devices to the core, ensuring that resource-constrained IoT devices are not burdened with intensive processing tasks. DMF is validated using data from a real-world IoT deployment in Madrid, demonstrating the effectiveness of the proposed system in accurately estimating traffic, environmental, and pollution metrics from a reduced set of sensors. The proposed solution offers a scalable, efficient mechanism for managing urban IoT networks, while addressing issues of sensor failure and privacy concerns.
\end{abstract}


%
\IEEEpeerreviewmaketitle

\section{Introduction}
\label{intro}

The Internet of Things (IoT) and sensor networks are critical components in the design and operation of smart cities. These systems enable real-time monitoring, data collection, and dynamic management of urban environments. Sensors distributed across a city, such as those monitoring traffic flow, air quality, energy consumption, waste management, and public safety, generate vast amounts of data that inform the operation and well-being of the urban ecosystem. IoT infrastructure connects these sensors, facilitating seamless communication and data integration across city services like transportation, emergency response, and environmental management. These IoT-based smart city solutions enhance sustainability by optimizing resource usage, reducing energy consumption, and improving waste management practices, thus driving cities toward more efficient, responsive, and sustainable operations.

The efficient operation of smart cities requires extensive networks of IoT devices collecting data from various sensing modalities. For example, Madrid's traffic-monitoring infrastructure includes over 7,000 sensors deployed at more than 4,000 locations, generating around 145 million data points. Similar networks manage other modalities like pollution levels, meteorological conditions, noise pollution, and energy usage. Managing this large-scale data ecosystem poses several challenges. First, scalable approaches are required for efficient storage, processing, and transmission of the massive data volumes generated. Communication resources, particularly bandwidth for data transmission to cloud-based systems, are also a concern. In addition, the energy expenditure of resource-constrained IoT devices leads to increased carbon footprints, contrary to the sustainability goals of smart city initiatives. Finally, sensor network maintenance is costly, and solutions must address sensor failures effectively.

Previous research \cite{abbasian2020survey, arumugam2015ee, halder2019limca, chen2019layered, chowdhury2020adaptive, yang2023guidelines, alam2021error, he2019multi} has addressed data handling in sensor networks, focusing on communication bandwidth and energy optimization. Techniques like Singular Value Decomposition (SVD), Principal Component Analysis (PCA), and machine learning methods such as autoencoders have been applied for data compression, but these often require computational resources that far exceed the capabilities of resource-constrained IoT devices. These methods impose significant energy and processing burdens on sensor nodes and primarily address data management without offering scalable solutions for managing large sensor networks. The concept of synthetic sensing, as proposed in \cite{laput2017synthetic, laput2019sensing}, reduces isolated sensing modalities in smart home systems by leveraging a small number of sensors to infer secondary events. However, these systems still require considerable on-device computation and lack the granularity needed for real-time urban-scale applications.

This paper introduces a novel approach, Data-driven Modality Fusion (DMF), to address these challenges. By exploiting correlations between time-series data across multiple sensing modalities, the proposed framework enables high-granularity real-time estimation of various sensing parameters—such as traffic intensity, temperature, humidity, and noise levels—from a reduced set of pollution sensor data. The DMF approach shifts the computational burden away from edge devices to the core, thus reducing energy consumption, bandwidth usage, and deployment costs. This method also enhances sensor network robustness by enabling data recovery in case of sensor failure. Importantly, the framework avoids reliance on video or audio data, thereby mitigating privacy concerns. The effectiveness of DMF is demonstrated through empirical validation using data from a smart city deployment in Madrid.

The paper has the following specific scopes and contributions.

\begin{enumerate}
    \item A novel paradigm of Data-driven Modality Fusion is proposed for sensing cost reduction for efficient urban IoT network management.
    \item The framework is designed to achieve the multi-dimensional objectives of data reconstruction, network size reduction and energy-bandwidth usage management; without the involvement of any additional computation at the sensor nodes.
    \item The proposed concept is validated in a real world smart-city IoT network deployed in Madrid.
    \item A detailed characterization of the framework to analyze the trade-off between computation cost and performance is presented.
    \item A comprehensive study on data correlation and eigen decomposition analysis is provided to understand the feasibilty and limits of the proposed mechanism.
\end{enumerate}

The remainder of this paper is organized as follows. Section \ref{secii} provides a critical review of existing similar works and points out how this paper aims at solving the gaps. The system architecture, including the components of the smart city IoT network, on which the proposed concept is built, is presented in section \ref{seciii}. In section \ref{corr}, we explain the correlation and inter-dependency among data from different sensing modaliites, which is then leveraged to implement the Data-driven Modality Fusion framework. detailed in section \ref{proposal}. Performance of the proposed framework is reported in section \ref{results} and an eigen space data representation is presented in section \ref{eigen_rep} to understand and illustrate the feasibility of the proposed system, with the conclusions drawn in section \ref{conc}.








\section{Related Work}
\label{secii}

There are existing research works that address data handling in sensor networks, primarily from the standpoint of managing communication bandwidth costs and minimizing sensing energy expenditure. Several approaches have been proposed in the literature, focusing on data compression techniques. As outlined in \cite{ketshabetswe2021data}, these methods can be divided into two primary categories: local data compression and distributed data compression. In distributed data compression \cite{abbasian2020survey, arumugam2015ee, halder2019limca, chen2019layered}, data from multiple neighboring sensor nodes is aggregated to remove redundant information before transmission. This aggregation typically occurs at a central base station or a cluster head within the network. For this to be effective, spatial correlations must exist in the data collected by multiple sensor nodes. However, this assumption is not always valid for certain applications. In contrast, local data compression techniques rely on temporal correlations within individual sensor nodes to reduce the amount of data that needs to be transmitted \cite{cao2020multi, jarwan2019data}. The work in \cite{chen2019hierarchical} introduced a compression technique that utilizes both spatial and temporal correlations in sensor data. These methods compress the data at the sensor nodes, either in a lossless or lossy manner. Lossless compression methods, such as Huffman coding, run-length encoding, and other statistical techniques, have been developed in \cite{saidani2019new, gao2022application, ahmad2019lossless}. Lossy compression techniques, including Bayesian Predictive Coding and piecewise linear approximation, have been proposed in \cite{chen2020new, roy2022limited}, but they often suffer from irreversible data loss while also imposing a computational burden on sensor nodes.

Additionally, several research efforts focus on leveraging machine learning for data compression in resource-constrained sensor and IoT networks, aimed at improving energy efficiency. A much-explored methodology in these works is to encode or compress time-series data into relevant features, which are then transmitted to the cloud for further analysis. Common dimensionality reduction techniques such as Principal Component Analysis (PCA) \cite{chowdhury2020adaptive, yang2023guidelines, diwakaran2019cluster}, Singular Value Decomposition (SVD) \cite{alam2021error, he2019multi}, and Linear Discriminant Analysis (LDA) \cite{fabiyi2021folded} have been widely adopted for compressive sensing in wireless environments. In these methods, the optimal number of components to retain is determined by factors specific to the application, which adds complexity. Furthermore, these techniques require storing the original data matrix, presenting challenges for sensor nodes with limited memory and computational power. Other machine learning-based compression techniques, including autoencoders \cite{kuester2023convolutional, chen2020wsn} and reinforcement learning \cite{yun2021q}, have also been explored. Downstream task-driven semantic communication techniques \cite{feng2023data, wang2023semantic} have also been applied to wireless networks as a novel data compression strategy. 

The concept of Synthetic Sensing and General Purpose Sensing, as proposed in \cite{laput2017synthetic, laput2019sensing, zhang2018vibrosight}, offers an alternative approach by reducing the number of sensing modalities required for a smart home IoT system. This approach involves deploying a small number of sensors on a single Printed Circuit Board (PCB), capable of generating machine learning features to monitor multiple secondary events. In \cite{laput2017synthetic}, the authors demonstrate how a single complex sensor can indirectly monitor a large context, without direct instrumentation of objects. The paper also demonstrates the system's ability to virtualize raw sensor data into actionable feeds, whilst simultaneously lowering immediate privacy issues. Similarly, in paper\cite{laput2019sensing}, the authors explore the feasibility of sensing hand activities from commodity smartwatches. The work presented in \cite{zhang2018vibrosight} develops an approach to sense activities in a room using long-range laser vibrometry as an alternative to using sensing units relying camera or microphone. While this approach of synthetic/general-purpose sensing simplifies sensor deployment and reduces privacy issues, it still requires significant processing complexity at the sensor nodes. Furthermore, it does not support the fine-grained real-time estimation of multiple modalities that is often needed in smart city IoT networks, which demand more dynamic and scalable solutions.

However, in most of these approaches, the data processing required for compression takes place at the sensor nodes. This introduces significant challenges for resource-constrained IoT devices that lack the necessary hardware support, such as processing and storage capabilities, to execute computationally intensive learning algorithms. For instance, in the context of smart city IoT networks, where a large number of sensors are deployed, executing complex machine learning algorithms at the sensor nodes can lead to significant energy consumption, which poses a limitation from both processing and energy efficiency perspectives. Moreover, these methodologies primarily focus on data management and do not address the broader issue of managing the large number of sensors deployed in smart city IoT networks.

\label{seciii}
\begin{figure*}[h]
	\centering
	\includegraphics[width=16cm,height=10.0cm]{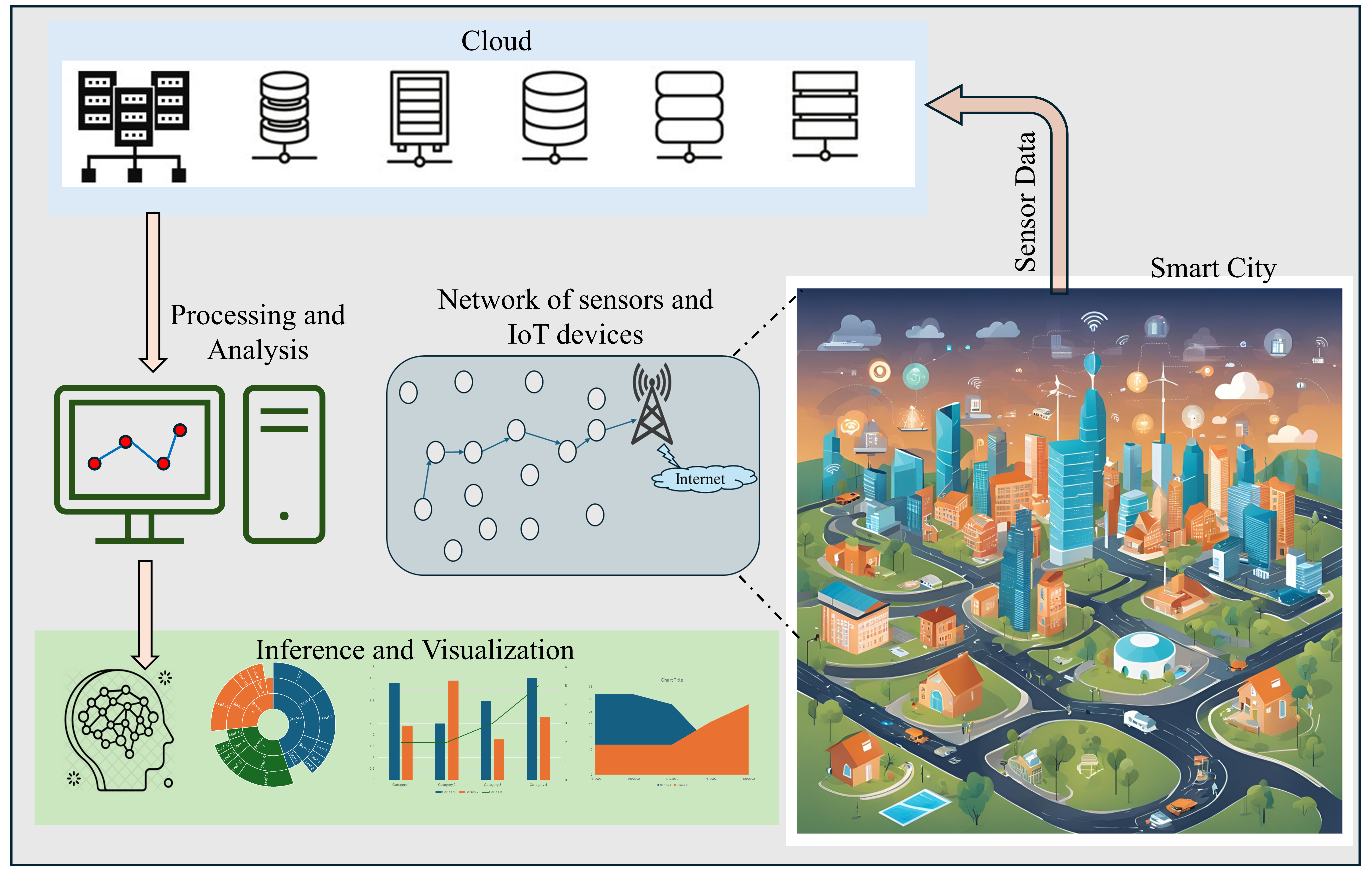}
	\caption{System model representing smart city IoT data collection, processing and analysis}
	\label{sys_dia}
\end{figure*}

\section{System Model}

In this work, we consider the IoT network deployed in the smart city of Madrid as the testbed for demonstrating the proposed concept of Data-driven Modality Fusion (DMF). As shown in Fig. \ref{sys_dia}, the city of Madrid has numerous sensors and IoT devices spread across the city for an efficient urban management assistance. There are sensors gathering spatio-temporal data for different modalities, such as, traffic intensity, air quality, noise pollution, meteorological data, such as, temperature, humidity, solar radiation, wind speed and direction, building energy consumption, to name a few. These sensors collect data at different sampling rate as per the application requirement, for example, traffic intensity data is updated every 15 minutes in real time, air quality and meteorological data is sampled at an hourly basis, while noise pollution level is recorded on a daily basis. 

Different state-of-the-art sensing techniques are employed for data collection. The most common technique used for traffic monitoring is by deploying magnetic loops embedded underneath the roads that detect the presence of vehicles by measuring changes in inductance as a car passes over the loop \cite{lana2016role}. In addition, there are several high-definition cameras, which are used for real-time vehicle classification and detection using image/video processing and AI techniques. There are Doppler radar and LIDAR sensors, as well, that are used for vehicle speed detection and tracking. This data enables the calculation of critical road usage metrics, including traffic flow, occupancy, load, level of service, and speed, which are essential for various traffic management applications. Similarly, the meteorological stations measure different environmental parameters using thermometers, anemometers, barometers, hygrometers, rain gauges and other specialized sensors deployed across the city. Madrid air quality system maintains 24 monitoring stations in the metropolitan area that measures the concentration of different particulate matter, such as, $PM_{2.5}, PM_{10}$, and other pollutant gases, including $CO_2, NO_X, SO_2, O_3$. The concentration of Ozone ($O_3$) is measured with ultraviolet absorption at 253.7 nm, nitrogen oxides using chemiluminescence analyzers, and $PM_{10}$ concentration is detected with a heated tapered oscillating microbalance (TEOM) \cite{nunez2019statistical}. An array of highly sensitive microphones deployed throughout the city is used to detect noise levels in decibels (dB) \cite{asensio2020changes}. Several frequency analyzers are used to detect specific frequencies of noise to classify types of noise pollution (traffic, industrial, etc.).

As for the IoT network, roadside devices, in general, transmit traffic data wirelessly to central nodes. Long-range, low-power networks, such as LPWAN (LoRa, NB-IoT), are commonly used to transmit traffic data to centralized servers in urban environments. WiFi and cellular networks (GPRS/4G/5G) facilitate real-time data upload for central processing and analytics. For remote or less accessible areas, meteorological data is sometimes collected via satellite and transferred to local or central servers. Centralized IoT platforms like FiWare or proprietary smart city dashboards are used to collect, process, and analyze the data in real time. Many smart cities, including Madrid, rely on cloud platforms for large-scale data storage, processing, and analytics. Open Data APIs are provided for public access, allowing users to query and integrate the data into applications or perform real-time analysis.

\begin{table*}[h] \centering
	\caption{Pearson Correlation Coefficient of pollutant concentration levels with other sensing modalities}
	\label{pear_corr_tab}
	\begin{tabular}{|l|l|l|l|l|l|l|l|l|}
		\hline
		\begin{tabular}[c]{@{}l@{}}Variable \\ (Pollutants)\end{tabular} & \multicolumn{1}{c|}{Traffic Intensity} & \multicolumn{1}{c|}{Noise Level} & \multicolumn{1}{c|}{Wind speed} & \multicolumn{1}{c|}{Wind direction} & \multicolumn{1}{c|}{Temperature} & \multicolumn{1}{c|}{Humidity} & \multicolumn{1}{c|}{Radiation} & \multicolumn{1}{c|}{Precipitation} \\ \hline
		$SO_2$                                                              & 0.207213                               & 0.102321                         & -0.150906                       & -0.105862                           & -0.411252                        & 0.105727                      & -0.068647                      & -0.055953                          \\ \hline
		$CO$                                                               & 0.22812                                & 0.136088                         & -0.476759                       & -0.220662                           & -0.327063                        & 0.144675                      & -0.220361                      & -0.052487                          \\ \hline
		$NO$                                                               & 0.21313                                & 0.227828                         & -0.35899                        & -0.117307                           & -0.338046                        & 0.251443                      & -0.149968                      & -0.029169                          \\ \hline
		$NO_2$                                                              & 0.204846                               & 0.20157                          & -0.581239                       & -0.194545                           & -0.25946                         & 0.137868                      & -0.287997                      & -0.040334                          \\ \hline
		$PM_{2.5}$                                                           & 0.063012                               & 0.079999                         & -0.316483                       & -0.126227                           & 0.092968                         & 0.115939                      & -0.014756                      & 0.049101                           \\ \hline
		$PM_{10}$                                                             & 0.151277                               & -0.018111                        & -0.249907                       & -0.054517                           & 0.209148                         & -0.106957                     & 0.058575                       & -0.02295                           \\ \hline
		$NO_X$                                                              & 0.226663                               & 0.229944                         & -0.482601                       & -0.159606                           & -0.331677                        & 0.223205                      & -0.220902                      & -0.036247                          \\ \hline
		$O_3$                                                               & 0.15589                                & -0.312573                        & 0.572664                        & 0.220748                            & 0.668691                         & -0.62378                      & 0.563749                       & 0.001184                           \\ \hline
		$C_6H_5CH_3$                                                              & 0.15291                                & 0.176781                        & -0.42011                        & -0.15519                            & -0.12279                         & 0.077028                      & -0.13574                       & -0.0383                           \\ \hline
		$C_6H_6$                                                              & 0.235303                               & 0.209266                       & -0.38368                        & -0.11549                            & -0.33473                         & 0.200116                      & -0.17579                       & -0.05026                           \\ \hline
		$C_8H_{10}$                                                              & 0.176519                               & 0.212734                        & -0.39863                        & -0.1489                            & -0.2303                         & 0.179424                      & -0.14129                       & -0.03538                           \\ \hline
	\end{tabular}
\end{table*}

\begin{table*}[h] \centering
	\caption{Spearman's Rank Correlation of pollutant concentration levels with other sensing modalities}
	\label{spear_corr_tab}
	\begin{tabular}{|l|l|l|l|l|l|l|l|l|}
		\hline
		\begin{tabular}[c]{@{}l@{}}Variable \\ (Pollutants)\end{tabular} & \multicolumn{1}{c|}{Traffic Intensity} & \multicolumn{1}{c|}{Noise Level} & \multicolumn{1}{c|}{Wind speed} & \multicolumn{1}{c|}{Wind direction} & \multicolumn{1}{c|}{Temperature} & \multicolumn{1}{c|}{Humidity} & \multicolumn{1}{c|}{Radiation} & \multicolumn{1}{c|}{Precipitation} \\ \hline
		$SO_2$                                                              & 0.168179                               & 0.112888                         & -0.114181                       & -0.052423                           & -0.482279                        & 0.156842                      & 0.007201                       & -0.089244                          \\ \hline
		$CO$                                                               & 0.237128                               & 0.114863                         & -0.575494                       & -0.269987                           & -0.333893                        & 0.107422                      & -0.159674                      & -0.131848                          \\ \hline
		$NO$                                                               & 0.383942                               & 0.33046                          & -0.451741                       & -0.112181                           & -0.34515                         & 0.222697                      & 0.113168                       & -0.107918                          \\ \hline
		$NO_2$                                                              & 0.199244                               & 0.207234                         & -0.700558                       & -0.182195                           & -0.287555                        & 0.191732                      & -0.216641                      & -0.088707                          \\ \hline
		$PM_{2.5}$                                                           & 0.059461                               & 0.051666                         & -0.40873                        & -0.128436                           & 0.15741                          & 0.05546                       & 0.019734                       & 0.02368                            \\ \hline
		$PM_{10}$                                                             & 0.171115                               & -0.001151                        & -0.353341                       & -0.077431                           & 0.245475                         & -0.131148                     & 0.072386                       & -0.083181                          \\ \hline
		$NO_X$                                                              & 0.238304                               & 0.24513                          & -0.674628                       & -0.176409                           & -0.332731                        & 0.231519                      & -0.146616                      & -0.0937                            \\ \hline
		$O_3$                                                               & 0.154482                               & -0.341753                        & 0.690794                        & 0.190997                            & 0.625396                         & -0.627298                     & 0.414143                       & -0.002126                          \\ \hline
		$C_6H_5CH_3$                                                              & 0.236438                                & 0.169601                        & -0.55247                        & -0.12885                            & -0.07171                         & 0030056                      & -0.08081                       & -0.13118                           \\ \hline
		$C_6H_6$                                                              & 0.355713                               & 0.235405                       & -0.4504                        & -0.05729                            & -0.394                         & 0.170514                      & -0.05026                       & -0.13099                           \\ \hline
		$C_8H_{10}$                                                              & 0.23915                               & 0.227786                        & -0.53931                        & -0.1598                            & -0.22131                         & 0.163096                      & -0.07729                       & -0.10035                           \\ \hline
	\end{tabular}
\end{table*}

The data collected across different sensing modalities, using the system architecture mentioned above, are then uploaded continuously to the cloud in real time. Sensing, processing, storing and uploading of such a huge amount of data consume a lot of networking resources, specifically, bandwidth, energy, memory, carbon footprint etc. In this paper, we develop a Data-driven Modality Fusion framework that allows reduction in the number of sensing modalities, without additional computation at the resource-constrained sensor nodes and yet ensuring the presence of all the high-granular information content. This is accomplished by leveraging the inherent correlation that exists among the data across various sensing modalities which is explained next in section \ref{corr}.




\section{Correlation among Sensing Modalities}
\label{corr}

\begin{figure*}[h]
	\centering
	\includegraphics[width=\textwidth, height=6.5cm]{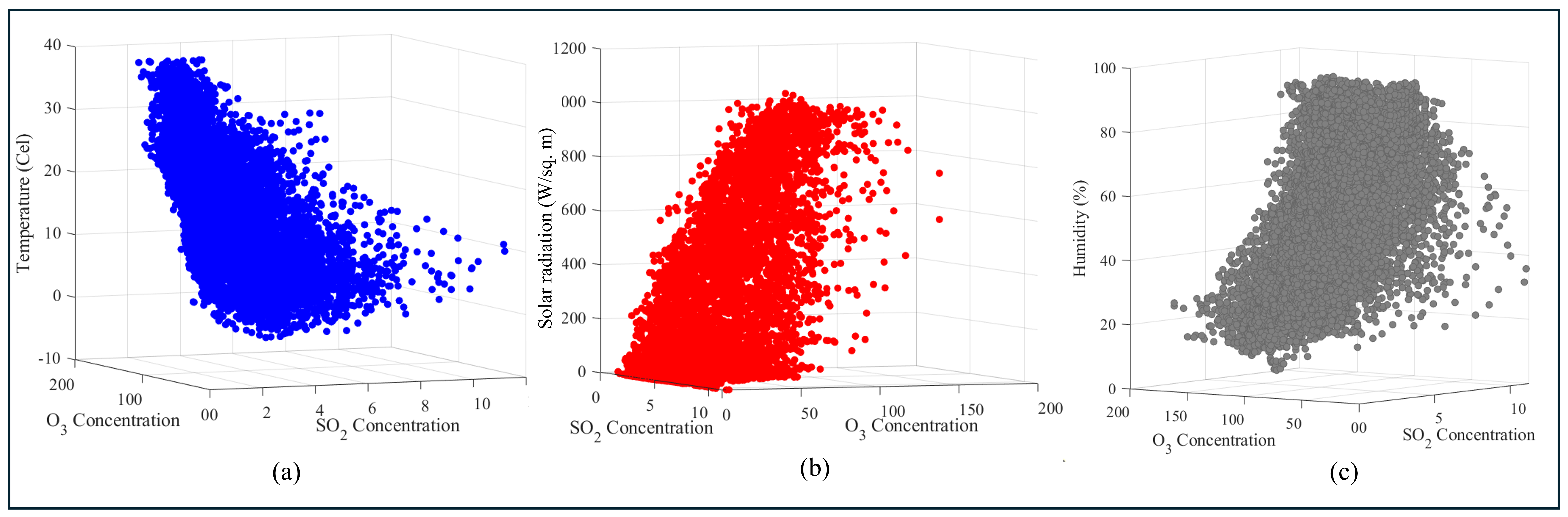}
	\caption{Inter-dependency among pollutant concentrations and meteorological sensing modalities}
	\label{corr_fig}
\end{figure*}

The proposed framework of Data-driven Modality Fusion (DMF) is based on the inter-dependency that exists among different sensing modalities. In the context of smart city IoT networks, the sensing modalities that we consider are time series of traffic intensity, meteorological data, noise levels, air quality and pollutant concentrations. In order to analyze the inter-dependency among these modalities, we consider two measures for correlation coefficients, viz, the standard Pearson Correlation Coefficient \cite{cohen2009pearson} and Spearman's Rank Correlation \cite{zar2005spearman}. While Pearson's correlation coefficient (denoted as $r$) measures the strength and direction of a linear relationship between two variables, while, Spearman’s rank correlation coefficient ($\rho$) is a non-parametric measure of the strength and direction of association between two ranked variables. It assesses how well the relationship between two variables can be described by a monotonic function, without assuming a linear relationship or normal distribution of the data. The expressions for computing Pearson's and Spearman’s rank correlation coefficients between two vectors $\{x_i\}$ and $\{y_i\}$, with corresponding sample means of $\bar{x}$ and $\bar{y}$ are given in Eqns. \ref{pearson} and \ref{spearman}, respectively, where $d_i$ is the difference between the ranks of $\{x_i\}$ and $\{y_i\}$.

\begin{equation}
	r = \frac{\sum_{i=1}^{n} (x_i - \bar{x})(y_i - \bar{y})}{\sqrt{\sum_{i=1}^{n} (x_i - \bar{x})^2 \sum_{i=1}^{n} (y_i - \bar{y})^2}}
	\label{pearson}
\end{equation}

\begin{equation}
	\rho = 1 - \frac{6 \sum_{i=1}^{n} d_i^2}{n(n^2 - 1)}
	\label{spearman}
\end{equation}

The correlation coefficients computed among the different sensing modalities, in the context of smart city IoT, are reported in Tables \ref{pear_corr_tab} and \ref{spear_corr_tab} respectively. Specifically, we compute  Pearson's and Spearman’s rank correlation coefficients of measured pollutant concentration levels with traffic intensity, noise levels and meteorological information for an entire year.

The reported correlation measures in the tables show the inter-dependency among various sensing modalities, especially for traffic intensity, noise level, wind speed, temperature and humidity with different pollutants' concentration levels. In a physical sense, this is nothing beyond expectation, since the increase in city traffic would lead to more noise and air pollution; and also affecting the weather and climatic conditions. Now the question that needs to be answered is if we can leverage this inter-dependency and correlation among data across different modalities to derive one from the other. With the aim of answering that question, we first consider the scenario of representing traffic, noise and meteorological information as a function of the pollutant concentrations ($SO_2, CO, NO,PM_{2.5},PM_{10},NO_X, O_3$, $C_6H_6$, $C_6H_5CH_3$, $C_8H_{10}$). To demonstrate the feasibility, we plot the year-wide meteorological data (temperature, radiation and humidity) as a function of $SO_2$ and $O_3$ concentrations, which is shown in Fig. \ref{corr_fig}. In order to simplify the 9-D representation to show the dependencies with all the pollutant concentrations, we consider only the pollutants $SO_2$ and $O_3$. From the figure, it can be observed that temperature, radiation and humidity can be  represented as close approximations of 2-D functions of $SO_2$ and $O_3$. The same logic can be applied to traffic and noise level sensing as well. The idea is that with the incorporation of all the pollutants' concentrations, the approximation becomes close to their true relationships. In other words, we should be able to model the relationships, given the sufficient number of feature vectors representing the pollutant concentrations.

\section{Data-driven Modality Fusion}
\label{proposal}

\begin{figure*}[h]
	\centering
	\includegraphics[width=\textwidth,height=10.0cm]{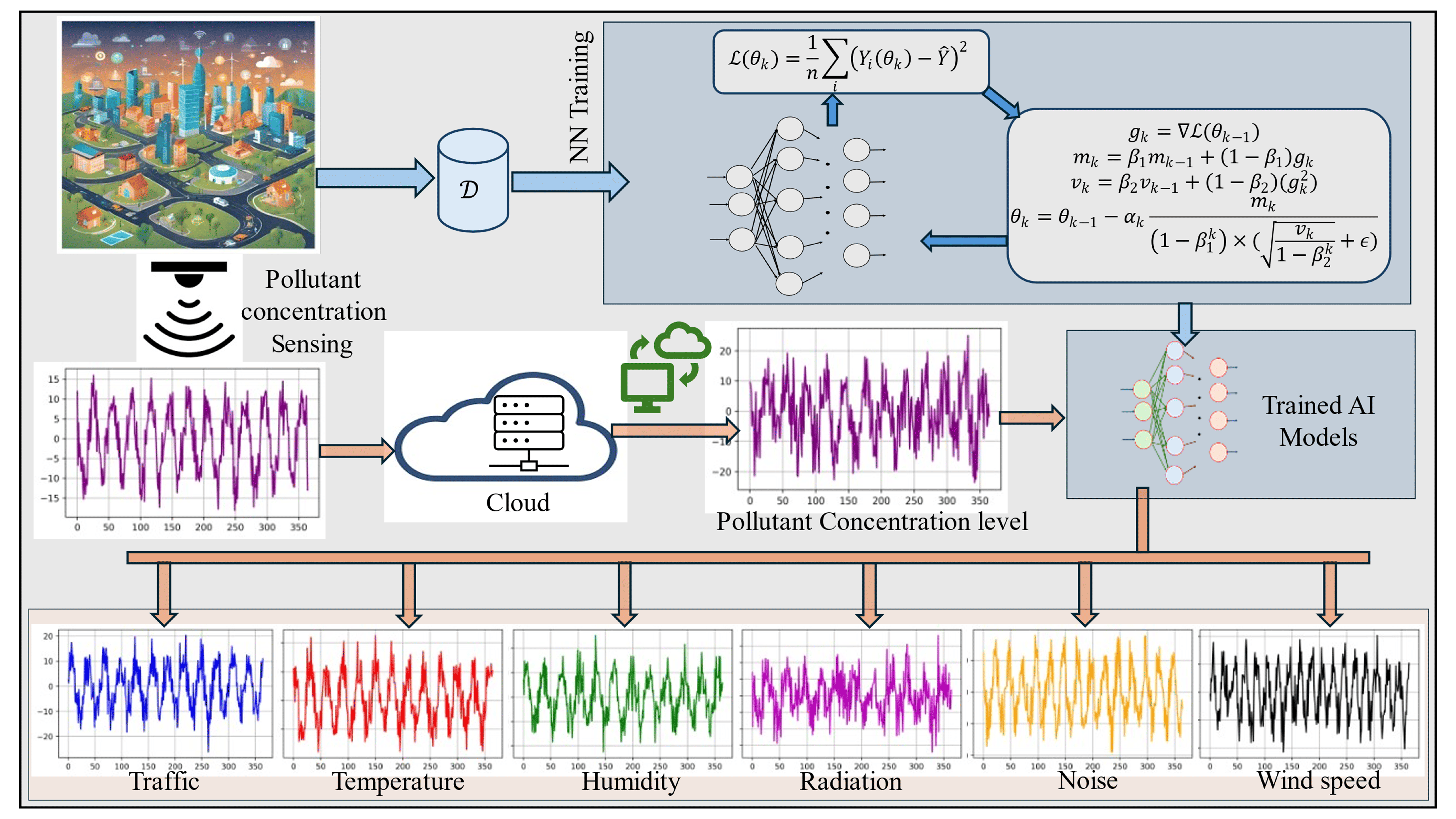}
	\caption{Data-driven Modality Fusion Framework.}
	\label{workflow}
\end{figure*}

Building on the existing inter-dependency among different sensing modalities, we propose the Data-driven Modality Fusion (DMF) that allows estimation of all other sensing modalities solely from the data collected using the pollutant concentration measuring sensors. The high-level representation of the DMF framework is presented in Fig. \ref{workflow}. Following are the main architectural changes in the proposed system as compared to the traditional approach in Fig. \ref{sys_dia}. On the IoT network side, the sensors for measuring traffic intensity, noise levels, solar radiation, temperature, humidity and wind speed are no longer required in the proposed approach. The IoT network only needs sensors for measuring pollutant concentrations. These readings are then uploaded to the cloud on a sample-by-sample basis. The receiver side, which is a central processor, downloads these readings and estimate the corresponding values of other sensing modalities, viz., traffic, temperature, humidity, solar radiation, wind speed and noise pollution. This is accomplished using trained deep learning models whose parameters are learnt using the existing data representing the relationship among all the sensing modalities.

The core concept here is to approximate the set of functions $\braket{f_i}, i \in \{$Traffic intensity, Noise levels, Temperature, Humidity, Wind speed, solar radiation$\}$, with the set of function predictors of $SO_2, CO, NO,PM_{2.5},PM_{10},NO_X, O_3$, $C_6H_6$, $C_6H_5CH_3$, $C_8H_{10}$ and any other temporal information. In this work, the above function approximations are accomplished by multi-variate regression models implemented using Deep Neural Network architectures. Two different approaches are explored for the implementation of the multi-variate regression models: Isolated Target Regressors (ITR) and Unified Target Regressor (UTR). In Isolated Target Regressors (ITR) approach, independent regression models are employed for each of the modalities that needs to be estimated from the pollutant concentrations. In other words, for estimating $\mathcal{M}$ number of modalities, we employ $\mathcal{M}$ regression models independently trained to find the relationship of each of them with the independent predictors. The Mean Square Error (MSE) loss for a model estimating modality $m$ (for a training batch of size $N$), that needs to be minimized, for this ITR approach is defined by Eqn. \ref{mae_itr}.
\begin{equation}
	\mathcal{L}(m; \theta_k) = \sum_{i=1}^{N}{(f_m(\mathbf{x}_i; \theta_k)-y_m)^2};  1\leq m \leq |\mathcal{M}|
	\label{mae_itr}
\end{equation}
Here, $\theta_k$ is the set of model parameters at instance $k$ and $f_m(\mathbf{x}; \theta_k)$ is the function approximated for the $m^{th}$ modality for the input predictor vector $\mathbf{x}$.

On the other hand, in the Unified Target Regressor (UTR) approach, all the all the modalities are estimated using a single model. The loss function in this case can be defined as:
\begin{equation}
	\mathcal{L}(\theta_k) = \sum_{i=1}^{N}{\sum_{m=1}^{\mathcal{M}}{(f_m(\mathbf{x}_i; \theta_k)-y_m)^2}}
	\label{mae_utr}
\end{equation}

Note that using the Unified Target Regressor (UTR) approach provides better scalability as it does not require separate model for estimation of each modality. However, there is a trade-off in terms of estimation performance and computation (at the central receiver) associated with that. The detailed evaluation of such trade-offs are presented in the results shown in section \ref{results}.

Training of the models in both these two approaches follows similar procedure. The models are trained based on already existing Madrid IoT sensors data collected over the year of 2023. The MSE losses defined in Eqns. \ref{mae_itr} and \ref{mae_utr} are minimized using the Adaptive Moment Estimation Algorithm that computes individual adaptive learning rates for different parameters from estimates of first and second moments of the gradients \cite{kingma2014adam}. The model parameter update rule at instance $k$ is given by Eqn. \ref{adam}, where $g_k, v_k$ and $m_k$ represent the loss gradient, first and second order moment estimates respectively.
\begin{subequations}
	\label{adam}
	\begin{align}
		g_k = \nabla \mathcal{L}(\theta_{k-1})\\
		m_k = \beta_1 m_{k-1}+(1-\beta_1)g_k\\
		v_k = \beta_2 v_{k-1}+(1-\beta_2)(g_k)^2\\
		\theta_{k}=\theta_{k-1}-\alpha_k \times \frac{m_k}{(1-\beta_1^k)\times (\sqrt{\frac{v_k}{1-\beta_2^k}}+\epsilon)}
	\end{align}
\end{subequations}

The trained models, when deployed, are capable of synthesizing sensor measurements from different modalities, apart from the one that is being deployed in the smart city IoT network. The following are the key advantages that are achieved using the DMF architecture. First, the number of sensors in the smart city IoT network is drastically reduced. This framework allows collection of all relevant real-time data at the central processor from the pollutant concentration measurements. This reduces the sensing cost by allowing users to get rid of expensive traffic sensors, microphone, camera, along with other sensors for measuring solar radiation, temperature, wind speed, and humidity. Second, because of the reduction of sensing data, the bandwidth required for transmitting data to the cloud and receiving the data at the receiver is reduced. This also saves the communication energy expenses. Third, due to reduction of sensing data, the energy spent by the IoT network on sensing and transmission is reduced.

\begin{figure*}[h]
	\centering
	\includegraphics[width=\textwidth, height=7.5cm]{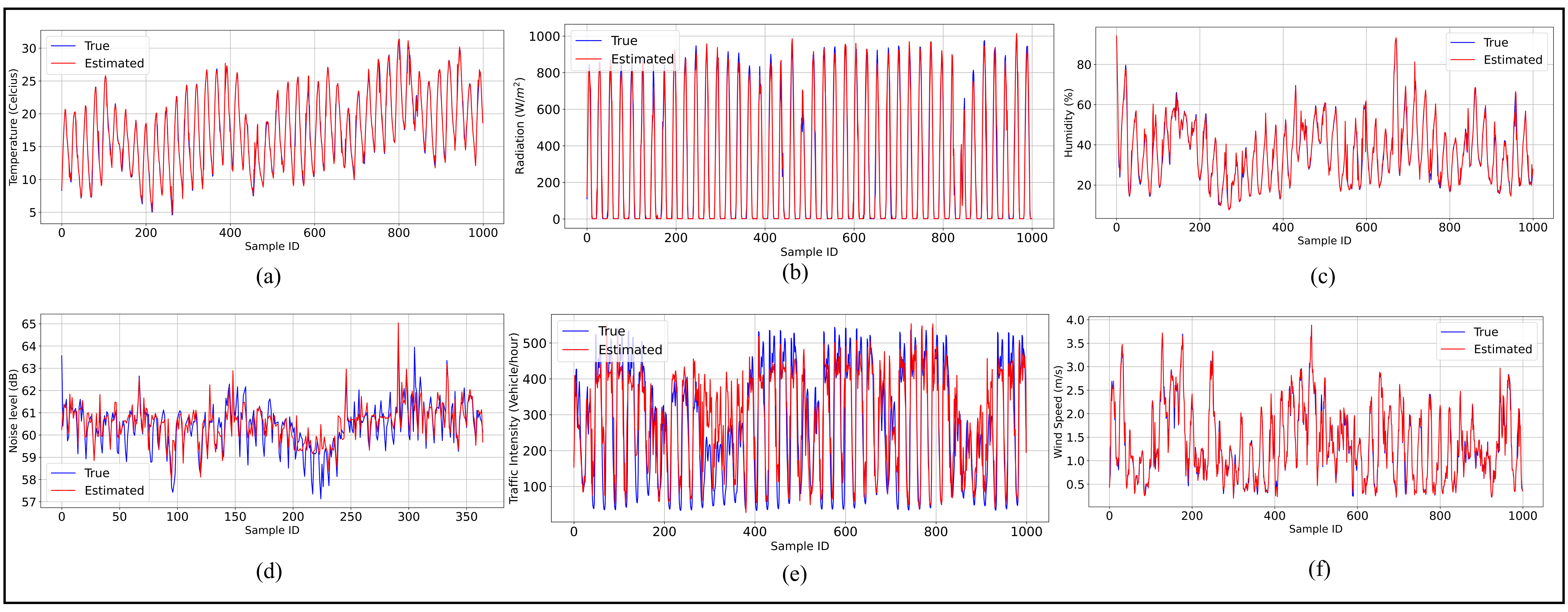}
	\caption{Comparison between the original and estimated time-series for different sensing modalities}
	\label{time_series}
\end{figure*}

It is noteworthy that the sampling rate of the synthesized modalities are bottlenecked by the data sampling and upload frequency of the physically deployed sensor network. Another point to be noted here is that choice of the predictor variable for the multivariate regression models is not unique. In other words, instead of synthesizing other sensing modalities from the pollutant concentrations, meteorological sensor data could be used for estimating traffic, noise or air quality index measurements. The primary reason for using pollutant concentrations as the prediction variable comes from their high correlation coefficients with the other modalities. Nevertheless, multiple such modalities could also be used as the regression predictor variables, which would allow to trade some bandwidth-energy overhead with estimation performance. Moreover, the reduction in sensing cost, energy and bandwidth usage is achieved at the expense of computation complexity at the central processor. Usually, these processors have high computational complexity and are externally powered, which allow them to execute these multiple complex deep learning architectures. Nevertheless, measures can be taken to reduce the complexity of the learning models at the receiver. One such approach relies on the trade-off between computational complexity and number of sensing parameters for a given prediction performance. In other words, complexity at the receiver (central processor) can be reduced by increasing the number of sensors at the transmitter (IoT device). All these scenarios are explored in the results presented in section \ref{results}.

\section{Performance Evaluation}
\label{results}

The system of Data-driven Modality Fusion (DMF) was evaluated using the data collected from the IoT network deployed in the city of Madrid\footnote{https://datos.madrid.es/portal/site/egob}. For demonstrating the working of the proposed architecture, time-series data recorded over the year of 2023, for multiple sensing modalities detailed above, was used for training the learning models. Details on the model hyperparameters are tabulated in Table \ref{model_param}. Performance is evaluated using Mean Absolute Error (MAE) between the true and estimated readings for each inferred modality values.  Since the actual value range for each sensing modality is different from one another, we normalize the MAE for a fair comparison of estimation performance across the different modalities. Normalized MAE (NMAE) for  modality $m$ is computed by min-max normalizing the MAE of $m$ with its range.

\begin{table}[]\centering
	\caption{Learning Model Baseline Hyperparameters}
	\label{model_param}
	\renewcommand{\arraystretch}{2}
	\begin{tabular}{|c|c|}
		\hline
		\textbf{Parameter}      & \textbf{Value} \\ \hline 
		Training Loss           & MSE            \\ \hline 
		Hidden Layer Activation & ReLU           \\ \hline 
		Learning Rate           & 0.001          \\ \hline 
		Batch Size              & 64             \\ \hline 
		Train-Validation Split        & 80:20          \\ \hline
		Stopping Criteria       & \makecell{$\nabla \mathcal{L}_{val}(\theta_{k-i}) \geq 0$ \\ $\forall i\ni (k-500)<i\leq k $}             \\ \hline
	\end{tabular}
\end{table}

Fig. \ref{time_series} shows a comparison of the estimated time-series value and true sensor readings for each of the six sensing parameters: traffic intensity, solar radiation, wind speed, noise pollution levels, temperature, and humidity. The estimation is done using the multivariate regression models described in section \ref{proposal}. For this set of experiments, Isolated Target Regressors (ITR) were used, where six independent models were trained for synthesizing each modality of the smart city IoT network. The figure demonstrates the perceivable similarity and correlation between the true and the estimated signal for the sensing parameters. It can be observed that the estimated signals by the learning models at the central server have similar amplitude and phase as the original signals sensed by the respective sensors. Moreover, since the prediction is done on a sample-by-sample basis (that is at the sensing sampling rate), there is no lag in the estimated signal, as is evident from the figure. 
\begin{figure}[h]
	\centering
	\includegraphics[width=3.3in]{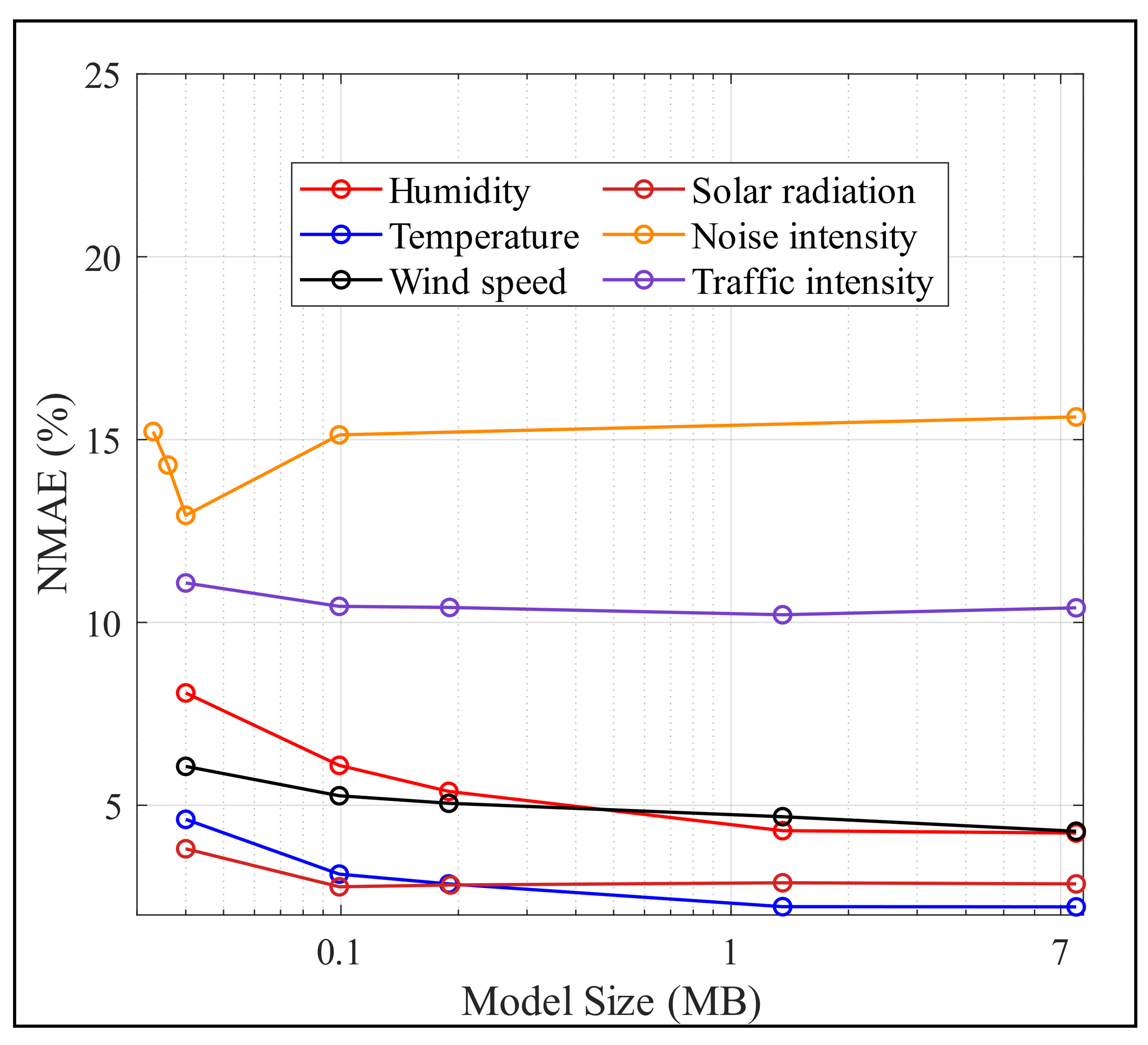}
	\caption{DMF Performance with ITR for different model complexity}
	\label{nmae_all_itr}
\end{figure}

Fig. \ref{nmae_all_itr} summarizes the DMF performance for different sensing modalities estimation with varying model complexity at the receiver. It is observed from the figure, in general, that estimation error goes down with increase in model size and then goes to a point of diminishing return beyond a certain complexity. This is because, an increase in model parameters provides higher degrees of freedom for approximating the non-linear relationship of the sensing modality with the pollutant concentration. Another point to note here is that estimation of noise levels has a higher error as compared to that of the other modalities. In addition, the error goes up with an increase in the model size beyond a certain point. The reason behind this is the low sampling rate of noise intensity values (sampled on a daily basis), and employing more complex models in such scenarios result in overfitting issues due to curse of dimensionality. This could be ameliorated by using multiple years of training data for noise level estimation. Nevertheless, the figure demonstrates that humidity, temperature, wind speed, solar radiation, noise and traffic intensity readings can be estimated with NMAE of 4.24, 2.22, 4.29, 2.77, 12.93 and 10.21$\%$ respectively. In other words, all these sensing modalities can be estimated solely from the pollution sensor data with an error bounded by 12.93$\%$.

\begin{figure}[h]
	\centering
	\includegraphics[width=3.3in]{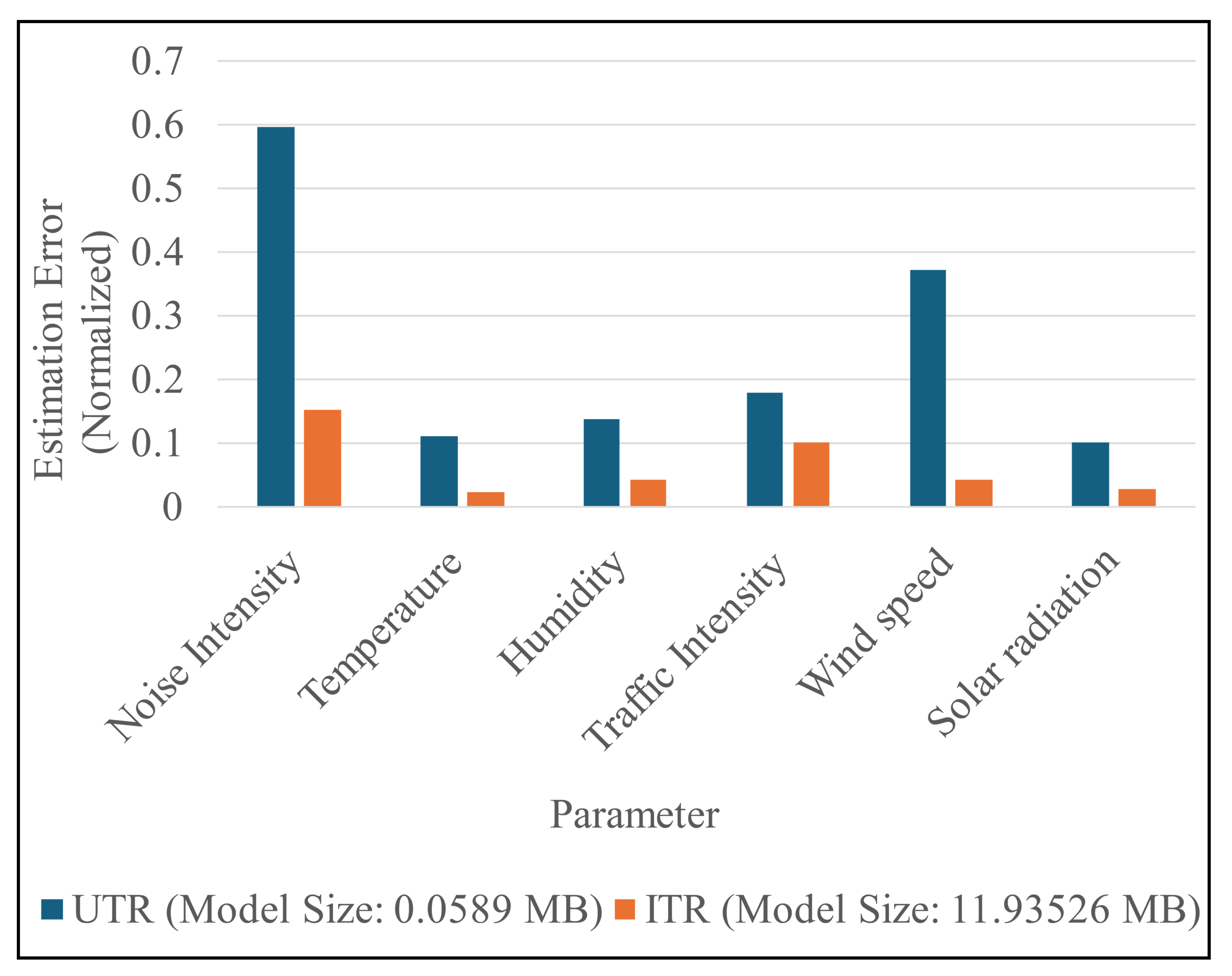}
	\caption{Performance comparison for ITR and UTR approaches}
	\label{utr_itr}
\end{figure}

The comparison between the two different approaches: Isolated Target Regressors (ITR) and Unified Target Regressor (ITR) is presented in Fig. \ref{utr_itr}. As observed in the figure, for estimation of all the modalities, ITR has a better performance than UTR. Especially noise intensity estimation using single model has NMAE that is four-orders of magnitude higher than that of using isolated models. The reason for this behavior can be attributed to the very diverse dynamic range of the parameters estimated. For example, wind speed is bounded in the range of [0.41, 3.41] m/s, whereas traffic intensity values fall in the range of [144, 445] counts per hour. This wide variation in the dynamic range creates problems in training a generalized regression model for estimating all the parameters simultaneously. However, note that in the ITR approach, the model complexity at the receiver is significantly higher than that in UTR approach, owing to the presence of several independent models operating parallely in the ITR execution. For the experiments performed, the model parameter counts in ITR was $10^6$ (i.e., 12 MB), whereas, in UTR, there are only $15K$ parameters (i.e., 0.06 MB), which is $\approx 67$ times lower than that of ITR. This trade-off allows the user to choose the balance between performance and receiver complexity in scenarios of resource constraints at the central processor.

\begin{figure}[h]
	\centering
	\includegraphics[width=3.1in]{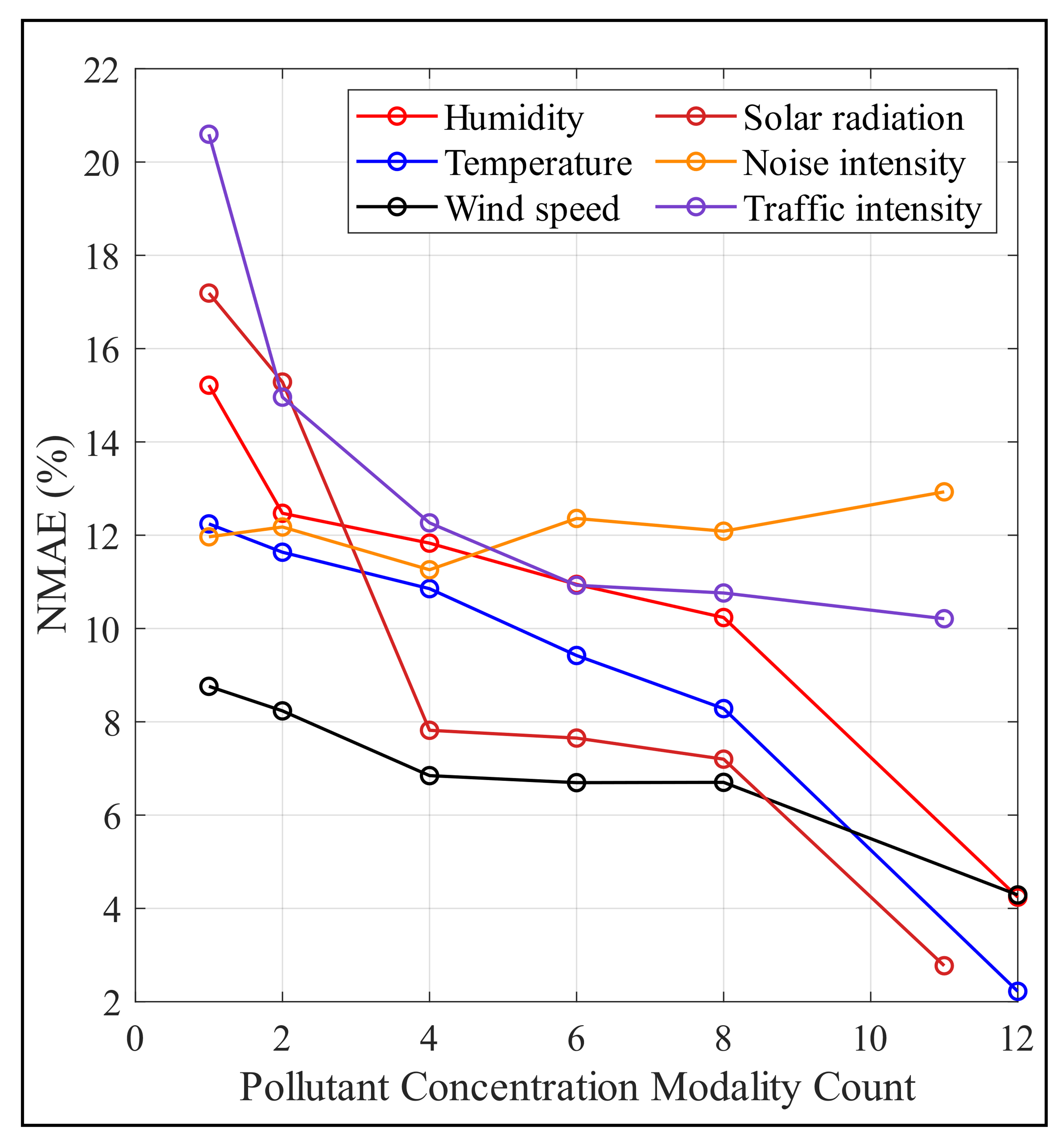}
	\caption{DMF Performance for varying sensing modality count}
	\label{nmae_poll_conc}
\end{figure}

In order to understand the effect of the number of  pollutant concentration measuring sensing modalities on DMF performance, we experiment with varying the number of pollution sensors and report the respective NMAE values for each estimated parameter in Fig. \ref{nmae_poll_conc}. Choosing the right set of pollution concentration measuring sensors is not trivial, since selecting different kinds of sensing modalities, even for the same number of modality count would affect model performance. Here, we select the modalities based on the correlation study presented in Tables \ref{pear_corr_tab} and \ref{spear_corr_tab}. The predictor sensing modalities are sorted based on their Pearman's correlation coefficient values with the modalities to be estimated. The top `$m$' modalities are then picked for training the DMF models at the receiver. One straight-forward observation from Fig. \ref{nmae_poll_conc} is the reduction of estimation error with an increase in the modalities used. This is because of the availability of higher-dimensional feature space for training the multi-variate regression models for a larger set of predictor modalities. The less sensitivity for noise level estimation can be explained by the low sampling rate of deployed microphone sensors, as explained above. Also, the meteorological estimation error drops significantly with the addition of temporal data as the $12^{th}$ modality.

\begin{figure}[h]
	\centering
	\includegraphics[width=3.1in]{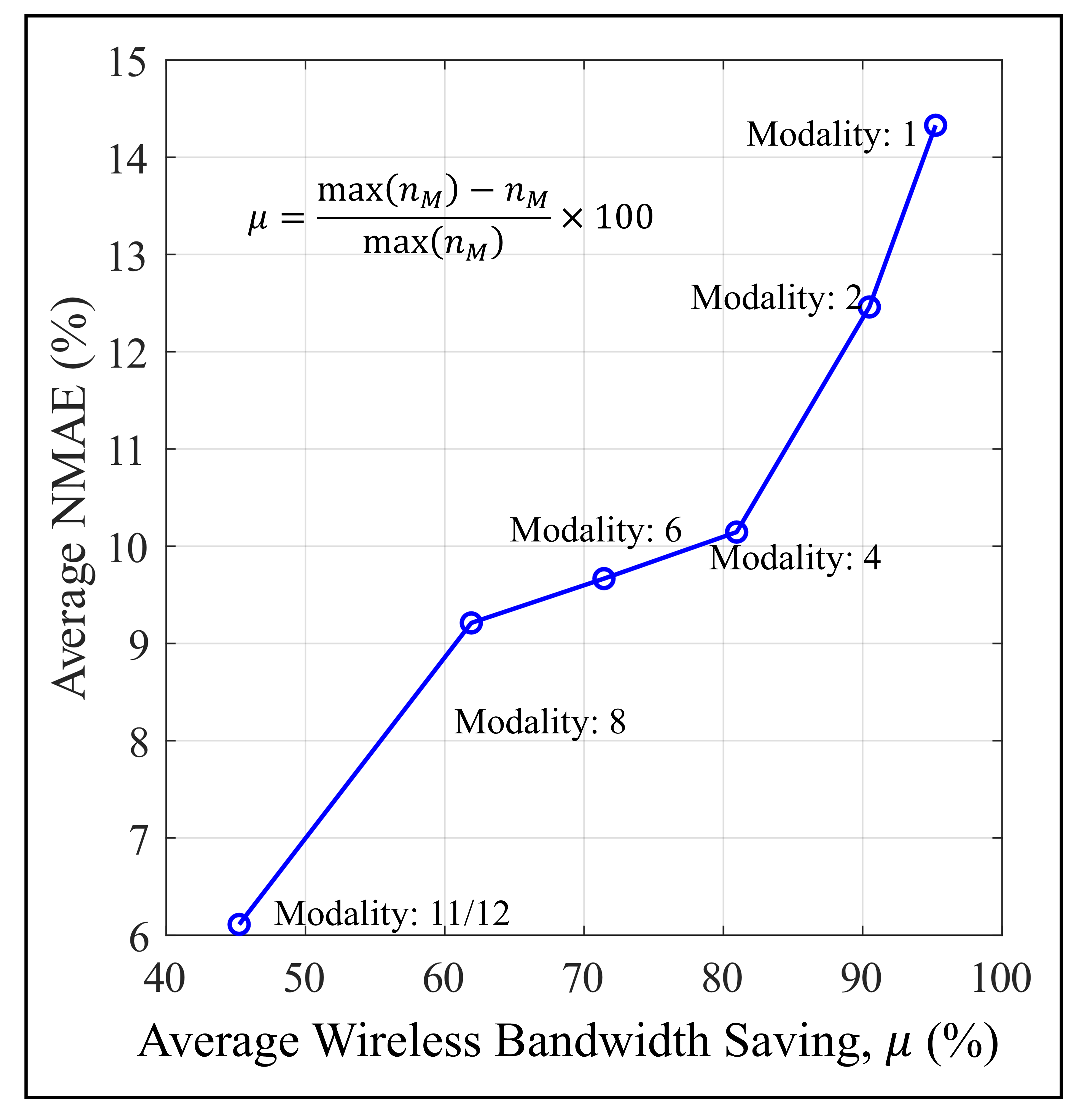}
	\caption{Trade-off between bandwidth usage and DMF performance ($n_M$: number of sensing modalites)}
	\label{nmae_bw}
\end{figure}

The key take-away from Fig. \ref{nmae_poll_conc} is that physical sensor counts for the smart city IoT network can be reduced even more by smart choice of predictor modalities. This helps to reduce data redundancy and helps sensor energy management and bandwidth usage reduction. The trade-off between bandwidth usage and DMF performance is represented in Fig. \ref{nmae_bw}, controlled by the physical sensor modality count. This trade-off would allow network operator or city management team to find the right balance between DMF estimation performance requirement and network bandwidth usage, based on application necessities.

\section{DMF Feasibility and Limits: Eigen Space Representation}
\label{eigen_rep}

The correlation study presented in section \ref{corr} helps in understanding the inherent relationships existing among different sensing modalities and serves as a guideline for choosing the smallest subset of modalities with an acceptable DMF performance. However, correlation simply measures pairwise linear relationships between variables without uncovering latent structures or global patterns. In addition, it is not entirely evident as to what is the relationship of correlation coefficients with the DMF performance, from the perspective of both estimation accuracy and network bandwidth usage. With the objective of providing an initial step towards understanding the feasibility of DMF from data analytics perspective, we provide an Eigen space representation of different sensing modalities discussed earlier. The motivation behind using Eigen space representation is that for a given dataset, Eigen vectors represent the direction along which it possesses the maximum variance.

For a given sensing modality $m$, we compute the Eigen vector $\mathbf{v}_m^*$ with the largest Eigen value $\lambda_m^*=max_p{\lambda_m(p)}$. This Eigen vector corresponding the largest Eigen value $\lambda_m^*$ can be derived using Eqn. \ref{eigen_eqn} 

\begin{equation}
	\mathbf{v}_m^*=argmax_{v_m}{(\mathbf{v}_m^TC_m\mathbf{v}_m)}
	\label{eigen_eqn}
\end{equation}

Here $C_m$ is the covariance matrix of the normalized dataset ($\mathcal{D}_m$), related to the $m^{th}$ modality, with $n_{\mathcal{D}_m}$ samples, which can be computed as follows.

\begin{equation}
	C_m = \frac{1}{n_{\mathcal{D}_m}-1}(\mathcal{D}_m-\mathbb{E}[\mathcal{D}_m])^T(\mathcal{D}_m-\mathbb{E}[\mathcal{D}_m])^T
	\label{covariance_eqn}
\end{equation}

\begin{figure}[h]
	\centering
	\includegraphics[width=3.3in]{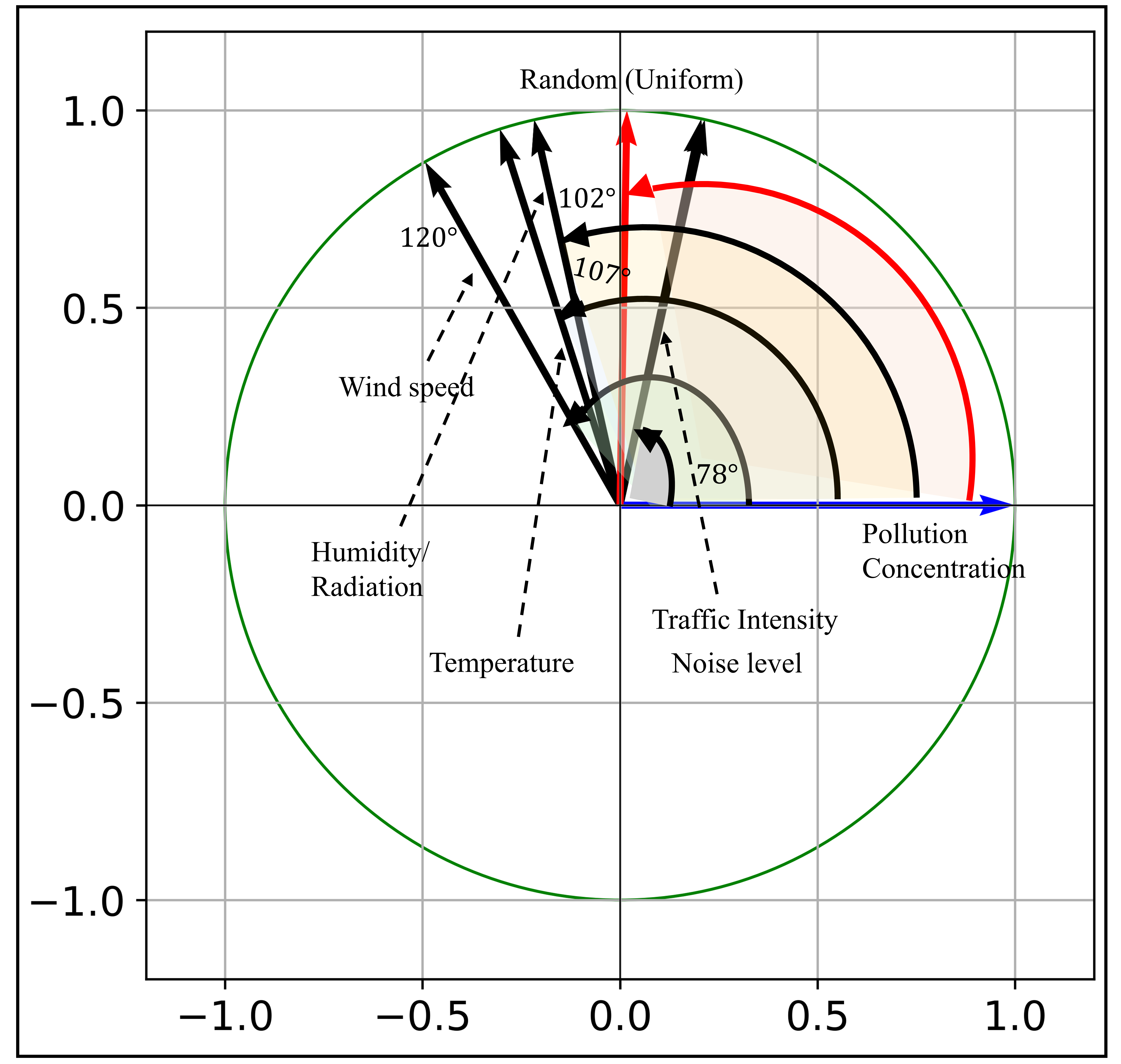}
	\caption{Representation of different sensing modalities in Eigen Space}
	\label{eigen_fig}
\end{figure}

Figure \ref{eigen_fig} presents the principal eigenvectors of various sensing modalities related to the smart city IoT network considered in this study. The angular distances between these eigenvectors and the eigenvector of pollutant concentration levels are also depicted. In this analysis, we consider data from eleven pollutants, including temporal information, to compute the principal eigenvectors.

As shown in Figure \ref{eigen_fig}, meteorological, traffic, and noise intensity data exhibit non-orthogonal relationships with pollutant concentration levels. Specifically, the traffic and noise intensity data form acute angles with the pollutant concentration eigenvector, indicating positive correlation, whereas the meteorological parameters form obtuse angles, indicating negative correlation. This observation is consistent with the correlation results in Tables \ref{pear_corr_tab} and \ref{spear_corr_tab}. Furthermore, we plot the eigen vector representing a dataset with random values drawn from a uniform distribution. It can be seen that the resulting eigen vector is orthogonal to the pollutant concentration, implicating least similarity and hence eliminating the scope of its synthesis from the pollutant concentration levels.

The representation of different modalities in the eigen space provides a direction to the feasibility of DMF among sensing modalities. However, more sophisticated measures are required that can provide a relationship between DMF accuracy and the angular distances in the Eigen space. This analysis represents an important first step towards understanding the potential and limitations of DMF in urban IoT networks.

\section{Conclusions}
\label{conc}
This paper presents the Data-driven Modality Fusion (DMF) paradigm as a promising solution for addressing the challenges associated with managing large-scale smart city IoT networks. By utilizing correlations between data streams from different sensing modalities, DMF reduces the need for numerous physical sensors while maintaining high accuracy in modality estimation. The empirical evaluation using real-world data from Madrid’s smart city deployment demonstrates that DMF achieves competitive performance in reconstructing key sensing parameters with minimal sensor infrastructure. The results reveal a trade-off between model complexity and estimation accuracy, allowing for flexible configuration based on the resource constraints at the network core and the application’s performance requirements.

DMF also provides a practical solution for reducing communication bandwidth usage and IoT device energy consumption, thereby aligning with the sustainability goals of smart city initiatives. The ability to estimate critical sensing modalities in real-time, even in the event of sensor failure, enhances the reliability and robustness of urban IoT systems. Furthermore, by offloading computation to the network core, the framework alleviates the burden on edge devices, making it particularly suitable for resource-constrained IoT networks.

Overall, DMF offers a scalable and effective method for managing smart city IoT networks, paving the way for more sustainable, resilient, and efficient urban environments. Future work can explore extending the DMF approach to incorporate spatial correlation among same-modality sensors for further efficiency gains. In addition, defining strict bounds and limits for DMF feasibility is also another direction for extension of this research.

\ifCLASSOPTIONcaptionsoff
  \newpage
\fi

\bibliographystyle{elsarticle-num}
\bibliography{Reference}

\end{document}